\documentclass[11pt,a4paper]{article}
\usepackage{jcappub}
\usepackage{euscript,epsfig,amsmath,amssymb}
\usepackage{amsfonts,latexsym}

\usepackage{graphicx}
\usepackage{color}

\newcommand{\be}[1]{\begin{equation}\label{#1}}
\newcommand{\ee}{\end{equation}}
\newcommand{\ba}[1]{\begin{eqnarray}\label{#1}}
\newcommand{\ea}{\end{eqnarray}}
\newcommand{\rf}[1]{(\ref{#1})}

\begin{document}

\title{{\bf Gurzadyan's Problem 5 and improvement of softenings\\
for cosmological simulations\\ using the PP method}}

\author{{\bf Maxim Eingorn}}

\affiliation{CREST and NASA Research Centers, North Carolina Central University,\\ Fayetteville st. 1801, Durham, North Carolina 27707, U.S.A.\\}

\emailAdd{maxim.eingorn@gmail.com}

\abstract{This paper is devoted to different modifications of two standard softenings of the gravitational attraction (namely the Plummer and Hernquist
softenings), which are commonly used in cosmological simulations based on the particle-particle (PP) method, and their comparison. It is demonstrated that some
of the proposed alternatives lead to almost the same accuracy as in the case of the pure Newtonian interaction, even despite the fact that the force resolution
is allowed to equal half the minimum interparticle distance. The revealed way of precision improvement gives an opportunity to succeed in solving Gurzadyan's
Problem 5 and bring modern computer codes up to a higher standard.}


\maketitle

\flushbottom

\

\section{Introduction}

\setcounter{equation}{0}

Among Gurzadyan's "10 key problems in stellar dynamics" \cite{Vahe}, any successful step towards solving the Problem 5 will exert powerful influence on the
state of affairs in stellar and galactic dynamics. The statement of this problem consists in creation of a computer code describing the $N$-body system with
the phase trajectory being close to that of the real physical system for long enough time scales. In view of primary importance of cosmological simulations in
analyzing the large scale structure formation and comparing results with predictions of different theories of the Universe evolution, the goal of the Problem 5
appears particularly important.

In this paper one such a step is proposed. Obviously, the higher precision can be achieved in $N$-body computer simulations, the more rigorous constraints can
be imposed on parameters of a concrete cosmological model. The well known PP method computing forces according to the Newton law of gravitation represents an
accurate $N$-body technique (see, e.g., \cite{Aarseth,Makino}). At the same time, it suffers from an evident shortcoming. Since the Newtonian gravitational
potential is singular at the particles' positions, softening is required in order to avoid divergences of forces when the interparticle separation distances
are very small. Introduction of softening ensures collisionless behavior of the system and simplifies numerical integration of its equations of motion
essentially. However, a high price to pay is noticeable reduction of precision. Below an attempt is made to modify two generally accepted softenings in such a
way that the accuracy of computer simulations becomes improved dramatically without any unjustified complication of the equations of motion or the integration
technique.

The paper is organized as follows. In the next section the equations of motion are written down and two standard softenings, namely, the Plummer and Hernquist
ones, are introduced. In the subsequent section various modifications are proposed and their efficiency is compared with respect to the same illustrative
example. Main results are discussed briefly in Conclusion.

\

\section{Plummer and Hernquist softenings}

According to the mechanical approach/nonrelativistic discrete cosmology, developed recently in a series of papers \cite{EZcosm1,EKZ2,EZcosm2} in the framework
of the conventional $\Lambda$CDM ($\Lambda$ Cold Dark Matter) model, the scalar cosmological perturbations in the Universe with flat spatial topology can be
described by the following perturbed FLRW metric in both nonrelativistic matter dominated and dark energy dominated eras:
\be{1} ds^2\approx a^2\left[(1+2\Phi)d\eta^2-(1-2\Phi)\delta_{\alpha\beta}dx^{\alpha}dx^{\beta}\right],\quad \alpha,\beta=1,2,3\, , \ee
where $a(\eta)$ is the scale factor depending on the conformal time $\eta$,
\be{2} \Phi(\eta,{\bf r})=\frac{\varphi({\bf r})}{c^2a(\eta)},\quad \triangle\varphi=4\pi G (\rho-\overline\rho)\, . \ee

Here ${\bf r}$ is the comoving radius-vector, $\triangle=\delta^{\alpha\beta}\partial^2/(\partial x^{\alpha}\partial x^{\beta})$ stands for the Laplace
operator, $G$ is the gravitational constant, and $\rho$ represents the rest mass density in the comoving coordinates $x^1\equiv x$, $x^2\equiv y$ and
$x^3\equiv z$. This quantity is time-independent within the adopted accuracy (both the nonrelativistic and weak field limits are applied), and $\overline\rho$
denotes its constant average value. Really, $\rho$ changes with the lapse of time in view of peculiar motion of cosmic bodies, however, the corresponding
velocities are small enough at the considered nonrelativistic matter dominated and dark energy dominated stages of the Universe evolution. Consequently, this
temporal change of $\rho$ may be disregarded when determining the gravitational potential from \rf{2}. In other words, it is determined by the positions of
cosmic bodies but not by their peculiar velocities, as it certainly should be in the nonrelativistic and weak field limits. Naturally, the introduced function
$\Phi$ \rf{2} satisfies the linearized Einstein equations for the metric \rf{1} within the adopted accuracy (see \cite{EZcosm1,EZcosm2}).

It is worth noting that in the considered flat spatial topology case the scale factor $a$ may be dimensionless, then the comoving coordinates $x^{\alpha}$ have
a dimension of length, and vice versa. In order to be specific, let us choose the first option, then the rest mass density $\rho$ is measured in its standard
units (namely mass$/$length$^3$).

In complete agreement with \cite{Peebles,gadget2,Landau2}, the following equations of motion, which describe dynamics of the $N$-body system experiencing both
the gravitational attraction between its constituents and the global cosmological expansion of the Universe, can be immediately derived from \rf{1} and
\rf{2}\footnote{The cutoff of the gravitational potential in the special relativity spirit introduced in \cite{EBV} in order to resolve the problem of nonzero
average values of first-order scalar perturbations is not considered here being irrelevant for the given investigation. Really, this cutoff is supposed to take
place at great distances of the order of the particle horizon and certainly does not affect dynamics on smaller scales described by the written down standard
equations of motion.}:
\be{3} \ddot {\bf R}_i -\frac{\ddot a}{a} {\bf R}_i=-G\sum_{j\neq i}\frac{m_j\left({\bf R}_i-{\bf R}_j\right)}{\left|{\bf R}_i-{\bf R}_j\right|^3}\, ,\ee
where ${\bf R}_i=a{\bf r}_i$ stands for the physical radius-vector of the $i$-th particle, and $m_i$ represents its mass. These equations are also commonly
used for simulations at astrophysical (i.e. non-cosmological) scales when the second term in the left hand side of \rf{3} is irrelevant and may be neglected.

Apparently, the right hand side of \rf{3} is a superposition of forces which originate from Newtonian gravitational potentials of single point-like particles.
If such a particle possessing the mass $m$ is located at the point ${\bf R}=0$, then its rest mass density in the physical coordinates
$\rho_{\mathrm{ph}}=m\delta({\bf R})$, and the potential of the produced gravitational field $\phi=-Gm/R$ is singular at the location point. In order to
suppress this singularity for reasons enumerated briefly in Introduction (namely, avoiding divergences of interparticle forces, ensuring collisionless dynamics
and simplifying numerical integration of equations of motion), let us consider two different models dealing with non-point-like gravitating masses. The density
and the potential of a single body in the Plummer model \cite{Plummer,Plummer2,PlumHern1,PlumHern2,PlumHern3}
\be{4} \rho_{\mathrm{ph}}^{(\mathrm{P})}=\frac{3m}{4\pi \varepsilon^3}\left(1+\frac{R^2}{\varepsilon^2}\right)^{-5/2},\quad
\phi^{(\mathrm{P})}=-\frac{Gm}{\sqrt{R^2+\varepsilon^2}}\, , \ee
while the same quantities in the Hernquist model \cite{PlumHern1,PlumHern2,PlumHern3,Hernquist}
\be{5} \rho_{\mathrm{ph}}^{(\mathrm{H})}=\frac{m\varepsilon}{2\pi R}\frac{1}{(R+\varepsilon)^3},\quad \phi^{(\mathrm{H})}=-\frac{Gm}{R+\varepsilon}\, , \ee
where $\varepsilon$ is the softening length/parameter (called the force resolution as well) typically amounting to a few percent of the mean interparticle
distance. While still dealing with point-like particles, one usually takes into account the gravitational interaction by means of $\phi^{(\mathrm{P})}$ or
$\phi^{(\mathrm{H})}$ in modern computer simulations based on the PP method. Thus, the force resolution $\varepsilon$ should not be attributed any real
physical sense representing just a mathematical trick eliminating singularity. Both Plummer $\phi^{(\mathrm{P})}$ and Hernquist $\phi^{(\mathrm{H})}$
potentials converge to the Newtonian one when $\varepsilon\rightarrow0$ (in this limit, as one can also easily verify, both densities in \rf{4} and \rf{5} tend
to the same expression $m\delta({\bf R})$ corresponding to a point-like massive particle, as expected). However, for any nonzero value of $\varepsilon$ the
attraction between every two bodies is changed with respect to the Newton law of gravitation at all separations. In particular, both analyzed potentials \rf{4}
and \rf{5} tend to zero as $-Gm/R$ when $R\rightarrow+\infty$ ($R\gg \varepsilon$), but in the opposite limit $R\rightarrow0$ ($R\ll \varepsilon$) they behave
as a constant $-Gm/\varepsilon$, so Newtonian singularity is absent. The next section is entirely devoted to controllable elimination of this defect of the
above-mentioned softenings.

\

\section{Illustration of accuracy improvement}

For illustration purposes let us consider a hyperbolic trajectory of a test particle with the mass $m$ in the gravitational field of the mass $M$ resting in
the origin of coordinates. This trajectory is given by the following functions \cite{Landau1} ($X$, $Y$ and $t$ denote Cartesian coordinates on the plane of
motion and time respectively):
\be{6} X=A(\epsilon-\cosh\xi),\quad Y=A\sqrt{\epsilon^2-1}\sinh\xi,\quad t=\sqrt{\frac{A^3}{GM}}(\epsilon\sinh\xi-\xi)\, ,\ee
where $\epsilon>1$ stands for the eccentricity, $\xi$ represents the varying parameter, and $A$ is the so-called semiaxis of a hyperbola, being interrelated
with the distance to perihelion $R_{\mathrm{min}}$: $A(\epsilon-1)=R_{\mathrm{min}}$. In what follows, the values $\epsilon=1.1$ and $\xi\in[0,0.15]$ are used.

The functions \rf{6} satisfy the equations of motion
\be{7} \frac{d^2X}{dt^2}=-GM\frac{X}{R^3},\quad \frac{d^2Y}{dt^2}=-GM\frac{Y}{R^3},\quad R=\sqrt{X^2+Y^2}\, .\ee

Introducing the normalized quantities
\be{8} \tilde X=\frac{X}{A}=\epsilon-\cosh\xi,\quad \tilde Y=\frac{Y}{A}=\sqrt{\epsilon^2-1}\sinh\xi,\quad \tilde
t=t\sqrt{\frac{GM}{A^3}}=\epsilon\sinh\xi-\xi\, ,\ee
one can rewrite \rf{7} in the form being more convenient for the subsequent numerical integration:
\be{9} \frac{d^2\tilde X}{d\tilde t^2}=-\frac{\tilde X}{\tilde R^3},\quad \frac{d^2\tilde Y}{d\tilde t^2}=-\frac{\tilde Y}{\tilde R^3},\quad \tilde
R=\sqrt{\tilde X^2+\tilde Y^2}\, .\ee

According to \rf{8}, if $\xi=0$, then $\tilde t=0$, $\tilde X=\epsilon-1$, $\tilde Y=0$, besides, $d\tilde X/d\tilde t=0$, $d\tilde Y/d\tilde
t=\sqrt{\epsilon^2-1}/(\epsilon-1)$. The enumerated values will serve as initial conditions hereinafter.

The exact solution \rf{8} is depicted on Fig. 1 (the black curve) together with the numerical solution of \rf{9} (red points). The leapfrog "drift-kick-drift"
numerical integration scheme is applied here and below with the fixed time step $\Delta\tilde t=0.0025$.

\begin{figure}[htbp]
\begin{center}\includegraphics[width=4.7in,height=2.98in]{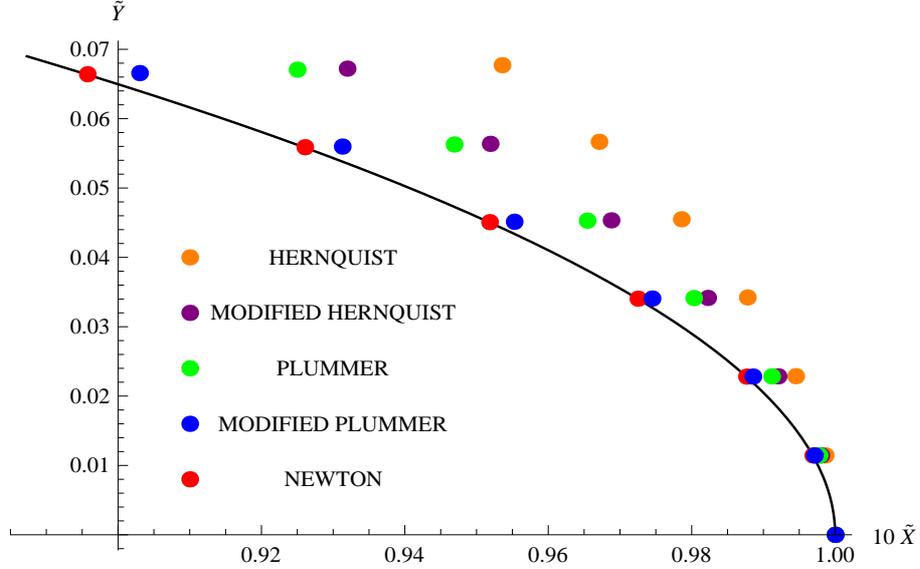}\end{center}
\caption{Trajectories for different potentials.}
\end{figure}

Orange points correspond to the modified equations of motion
\be{10} \frac{d^2\tilde X}{d\tilde t^2}=-\frac{\tilde X}{(\tilde R+\tilde\varepsilon)^2\tilde R},\quad \frac{d^2\tilde Y}{d\tilde t^2}=-\frac{\tilde Y}{(\tilde
R+\tilde\varepsilon)^2\tilde R}\, ,\ee
i.e. to the "Newton -- Hernquist" conversion
\be{11} \frac{1}{\tilde R}\rightarrow\frac{1}{\tilde R+\tilde\varepsilon}\ee
in the expression for the gravitational potential. The normalized softening length $\tilde\varepsilon$ everywhere equals $0.05$ (that is $\tilde\varepsilon$
amounts to $50\,\%$ of $\tilde R_{\mathrm{min}}$: $\tilde\varepsilon=\tilde R_{\mathrm{min}}/2$, meaning quite close approaching). Obviously, the orange points
lie rather far from the red ones. Consequently, the error is significant.

Further, one obtains purple points modifying the Hernquist potential:
\be{12} \frac{1}{\tilde R+\tilde\varepsilon}\rightarrow\frac{2}{\tilde R+\tilde \varepsilon}-\frac{1}{\tilde R+2\tilde \varepsilon}\, .\ee

The idea underlying this modification is simple: at each iteration one can make calculations using both $\tilde\varepsilon$ and $2\tilde\varepsilon$ softenings
and then interpolating results to the zero softening parameter. In other words, the expression in the right hand side of \rf{12} is constructed purposely in
such a way that for $\tilde R\gg\tilde\varepsilon$ its derivative with respect to $\tilde R$, being proportional to the gravitation force, behaves as
$-1/\tilde R^2+6/\tilde R^2\cdot(\tilde\varepsilon/\tilde R)^2$ up to the second order of smallness concerning the ratio $\tilde\varepsilon/\tilde R$. The term
of the first order is missing, therefore, the actual superposition of two Hernquist potentials with different softenings reduces the simulation error in
comparison with the previous case. Really, the purple points are noticeably closer to the red ones, then the orange points. However, precision is still low. Of
course, one can increase a number of terms in the superposition and apply a higher order interpolation, but introduction of each additional term requires more
computational time and, consequently, is not reasonable.

Green points correspond to the modified equations of motion
\be{13} \frac{d^2\tilde X}{d\tilde t^2}=-\frac{\tilde X}{(\tilde R^2+\tilde\varepsilon^2)^{3/2}},\quad \frac{d^2\tilde Y}{d\tilde t^2}=-\frac{\tilde Y}{(\tilde
R^2+\tilde\varepsilon^2)^{3/2}}\, ,\ee
i.e. to the "Newton -- Plummer" conversion
\be{14} \frac{1}{\tilde R}\rightarrow\frac{1}{\sqrt{\tilde R^2+\tilde\varepsilon^2}}\ee
in the expression for the gravitational potential. Precision is higher than in the previous case because the derivative of the expression in the right hand
side of \rf{14} with respect to $\tilde R$ for $\tilde R\gg\tilde\varepsilon$ behaves as $-1/\tilde R^2+1.5/\tilde R^2\cdot(\tilde\varepsilon/\tilde R)^2$, so
the deviation from the pure Newtonian behavior $-1/\tilde R^2$ is now four times smaller.

Finally, one gets blue points modifying the Plummer potential \cite{OhLin}:
\be{15} \frac{1}{\sqrt{\tilde R^2+\tilde\varepsilon^2}}\rightarrow\frac{1}{(\tilde R^4+\tilde\varepsilon^4)^{1/4}}\, .\ee

Now for $\tilde R\gg\tilde\varepsilon$ the deviation from the pure Newtonian behavior represents a quantity of the forth order of smallness concerning the
ratio $\tilde\varepsilon/\tilde R$. Consequently, precision is really high even despite the fact that the condition $\tilde\varepsilon=\tilde
R_{\mathrm{min}}/2$ holds true as before.

\section{Conclusion}

In this paper a promising opportunity of increasing the accuracy of computer $N$-body simulations based on the PP method is addressed. Namely, the inevitable
error arising from gravitational softening is reduced considerably by modifying the commonly used Plummer $\sim 1/\sqrt{R^2+\varepsilon^2}$ and Hernquist $\sim
1/(R+\varepsilon)$ potentials. In particular, the proposed $\sim 1/\left(R^n+\varepsilon^n\right)^{1/n}$ potential with $n>2$ gives better approximation since
for $R>\varepsilon$ the corresponding gravitation force differs from the standard Newtonian one in a small quantity $\sim(\varepsilon/R)^n$. This is
demonstrated explicitly for $n=4$ with the help of the concrete illustrative example of one particle moving along the hyperbolic trajectory in the softened
gravitational field of another one. The force resolution $\varepsilon$ is taken amounting to half the minimum separation distance, but despite this fact the
suggested alternative softening is characterized by much higher precision being much closer to the pure Newtonian picture than the standard ones.

Apparently, while improving numerical integration in the region $R>\varepsilon$ (where owing to this important inequality the expansion into series with
respect to the ratio $\varepsilon/R<1$ is allowed), the developed scheme still misrepresents the picture for $R\leqslant\varepsilon$ (where the above-mentioned
expansion is forbidden). However, if such close approachings happen seldom, this misrepresentation is not significant for the whole $N$-body system behavior
description. Thus, this scheme can really play an important role in astrophysical/cosmological modeling and solving the above-mentioned Problem 5. In other
words, the proposed modifications reducing simulation errors caused by softening can help to bring the phase trajectory of the $N$-body system in a
corresponding computer code much closer to that of the real physical one.

\

\section*{Acknowledgements}
My work was supported by NSF CREST award HRD-1345219 and NASA grant NNX09AV07A.
I would like to thank S.J.~Aarseth for valuable comments and the Referee for critical remarks which have considerably improved the presentation of the obtained
results.

\


\begin{thebibliography}{}

\bibitem{Vahe}
V.G. Gurzadyan, {\em 10 key problems in stellar dynamics: in retrospect}. In: "Ergodic Concepts in Stellar Dynamics", Eds. V.G. Gurzadyan, D. Pfenniger,
Lecture Notes in Physics {\bf 430}, p. 281-284 (Problems), p. 285-291 (Comments on key problems), Springer, 1994; (arXiv:astro-ph/1407.0398).


\bibitem{Aarseth}
S.J. Aarseth, {\em Gravitational N-body simulations: tools and algorithms}, Cambridge University Press, 2003.


\bibitem{Makino}
J. Makino, T. Fukushige, M. Koga and K. Namura, {\em GRAPE-6: massively-parallel special-purpose computer for astrophysical particle simulations}, PASJ {\bf
55} (2003) 1163.


\bibitem{EZcosm1}
M. Eingorn and A. Zhuk, {\em Hubble flows and gravitational potentials in observable Universe}, JCAP {\bf 09} (2012) 026; (arXiv:astro-ph/1205.2384).


\bibitem{EKZ2}
M. Eingorn, A. Kudinova and A. Zhuk, {\em Dynamics of astrophysical objects against the cosmological background}, JCAP {\bf 04} (2013) 010;
(arXiv:astro-ph/1211.4045).


\bibitem{EZcosm2}
M. Eingorn and A. Zhuk, {\em Remarks on mechanical approach to observable Universe}, JCAP {\bf 05} (2014) 024; (arXiv:astro-ph/1309.4924).


\bibitem{Peebles}
P.J.E. Peebles, {\em The large-scale structure of the Universe}, Princeton University Press, Princeton (1980).


\bibitem{gadget2}
V. Springel, {\em The cosmological simulation code GADGET-2}, MNRAS {\bf 364} (2005) 1105; (arXiv:astro-ph/0505010).


\bibitem{Landau2}
L.D. Landau and E.M. Lifshitz, {\it The Classical Theory of Fields, Fourth Edition: Volume 2 (Course of Theoretical Physics Series)}, Oxford Pergamon Press,
Oxford (2000).


\bibitem{EBV}
M. Eingorn, M. Brilenkov and B. Vlahovic, {\em Zero average values of cosmological perturbations as an indispensable condition for the theory and simulations}
(2014); (arXiv:astro-ph/1407.3244).




\bibitem{Plummer}
H.C. Plummer, {\em On the problem of distribution in globular star clusters}, MNRAS {\bf 71} (1911) 460.


\bibitem{Plummer2}
K. Dolag, S. Borgani, S. Schindler, A. Diaferio and A.M. Bykov, {\em Simulation techniques for cosmological simulations}, Space Science Reviews {\bf 134}
(2008) 229; (arXiv:astro-ph/0801.1023).


\bibitem{PlumHern1}
F. Iannuzzi and K. Dolag, {\em Adaptive gravitational softening in GADGET}, MNRAS {\bf 417} (2011) 2846; (arXiv:astro-ph/1107.2942).


\bibitem{PlumHern2}
J.E. Barnes, {\em Gravitational softening as a smoothing operation}, MNRAS {\bf 425} (2012) 1104; (arXiv:astro-ph/1205.2729).


\bibitem{PlumHern3}
B. R\"ottgers, T. Naab and L. Oser, {\em Stellar orbits in cosmological galaxy simulations: the connection to formation history and line-of-sight kinematics};
(arXiv:astro-ph/1406.6696).


\bibitem{Hernquist}
L. Hernquist, {\em An analytical model for spherical galaxies and bulges}, Astrophysical Journal {\bf 356} (1990) 359.


\bibitem{Landau1}
L.D. Landau and E.M. Lifshitz, {\it Mechanics, Third Edition: Volume 1 (Course of Theoretical Physics Series)}, Oxford Pergamon Press, Oxford (2000).




\bibitem{OhLin}
K.S. Oh, D.N.C. Lin and S.J. Aarseth, {\it On the tidal disruption of dwarf spheroidal galaxies around the galaxy}, Astrophysical Journal {\bf 442} (1995) 142.


\end{thebibliography}
\end{document}